\documentclass{lhep}       

\journal{LHEP}
\vol{xx}
\jyear{2018}
\pages{xxx} 

\received{xx January 2018}
\published{xx March 2018}

\def\be{\begin{equation}}
\def\ee{\end{equation}}
\def\bea{\begin{eqnarray}}
\def\eea{\end{eqnarray}}

\def\simlt{\stackrel{<}{{}_\sim}}
\def\simgt{\stackrel{>}{{}_\sim}}

\begin{document}

\title{Electroweak Baryogenesis and Higgs Physics}
\author{Carlos E.M. Wagner$^{1,2,3}$}
\address{$^1$Enrico Fermi Institute  and Kavli Institute for Cosmological Physics,}
\address{$^2$Physics Department, University of Chicago, Chicago, Illinois 60637, USA}
\address{$^3$HEP Division, Argonne National Laboratory, 9700 Cass Ave., Argonne, IL 60439}
\begin{abstract}
We discuss the constraints on the Higgs sector coming from the requirement of  the generation of the matter-antimatter asymmetry at 
the electroweak phase transition. These relate to both a strongly first order first transition, necessary for the preservation of
the generated baryon asymmetry, as well as CP-violation, necessary for its generation.  This scenario may lead to exotic decays of the Standard Model
like Higgs,  a deviation of the di-Higgs production cross section, or CP-violation in the Higgs sector. All these aspects are expected to be probed by the LHC as  well as by electric dipole moment experiments, among others. Further
phenomenological implications are discussed in this short  review.
\end{abstract}

\maketitle


\section{Introduction}
After the Higgs discovery at the LHC~\cite{ATLASHiggs, CMSHiggs} its properties including its mass and its couplings to the Standard Model (SM) particles have been studied in much detail~\cite{HiggsMass,ATLAScouplings,CMScouplings,ATLAS:2022vkf,CMS:2022dwd}. Those studies have shown that the Higgs boson has properties similar to the ones expected in the Standard Model. However,  only the couplings to the third generation quarks and charged leptons, and weak gauge
bosons are accurately known, with an uncertainty of the order of ten percent.   In addition, we know very little about the Higgs potential.  The SM potential
is fully specified by its minimum and the Higgs mass, which defines the quartic coupling of the Higgs,
\begin{equation}
V_{SM} = m_H^2 H^\dagger H + \frac{\lambda}{2} \left(H^\dagger H \right)^2.
\end{equation}
This implies $v^2 = - \frac{m_H^2}{\lambda}$,  with  the  neutral component ${\rm Re}[H^0]= \frac{1}{\sqrt{2} } (v+h) $,
 $v$ is the Higgs vacuum expectation value (vev) and $h$
is the normalized SM Higgs state, with mass $m_h^2 = \lambda v^2$.

Since the field $h$ and the vacuum expectation value act in a coherent way on the fermion fields, the fermion couplings are both diagonal and proportional to the fermion masses, $m_f$, namely
$g_{f_if_jh} = \frac{m_f}{v} \delta_{ij}$.
The coupling to the massive gauge bosons, on the other hand, is proportional to their mass squared, namely
$g_{hhV}  = \frac{m_V^2}{v}$.
The tree-level Higgs sector is therefore CP-conserving and all relevant CP-violating effects are in the charged  gauge boson couplings to quarks,
$g_{u_L^i d_L^jW^+} = \frac{g }{\sqrt{2}}V^{\rm CKM}_{ij}$,
with $V^{\rm CKM}$ being the CKM Unitary matrix, which contains  a phase, controlling CP-violation, $\delta_{CP}$.  
 All the above properties are relevant for the question of baryogenesis (see, for example Refs.~\cite{Cohen:1993nk,Trodden:1998ym,Riotto:1999yt,Morrissey:2012db,FileviezPerez:2022ypk}).
Assuming CPT conservation, Sakharov defined the conditions 
for a successful baryogenesis scenario~\cite{Sakharov:1967dj}, 
\begin{itemize}
\item Baryon Number Violation,
\item C and CP Violation,
\item Non-equilibrium processes.
\end{itemize}
Baryon Number Violation is present in the SM and is induced by so-called Sphaleron processes~\cite{Klinkhamer:1984di}, which are associated with the SM baryon and lepton number anomalous violating interaction, which at zero temperature are
vanishingly small, but at high temperatures are suppressed by a Boltzmann suppression factor~\cite{Arnold:1987mh,Arnold:1987zg},
\begin{equation}
\Gamma_{\rm Sph} \propto  T \exp(-E_{\rm Sph}/T).
\end{equation}
Here the Sphaleron energy $E_{\rm Sph} \sim  2 \pi v B/g$ and $B\simeq 4$~\cite{Klinkhamer:1984di}, implying a Sphaleron energy of the order of 10~TeV at zero temperature.

As mentioned before, CP violation processes are present in the SM, but the corresponding processes are suppressed by Jarlskog invariant  factors~\cite{Jarlskog:1985ht,Gavela:1994dt}, which combined with the  chiral suppression  associated with the small first and second generation quark masses, makes it impossible to get the observed baryon number density
\begin{equation}
\eta = \frac{n_B}{n_\gamma} \simeq 6 \times 10^{-10},
\end{equation}
where $n_B$ and $n_\gamma$ are the baryon and photon densities.  Hence, new CP-violating sources beyond the SM ones
are necessary for  baryogenesis.

\begin{figure}
\centering
\includegraphics[width=6cm,clip]{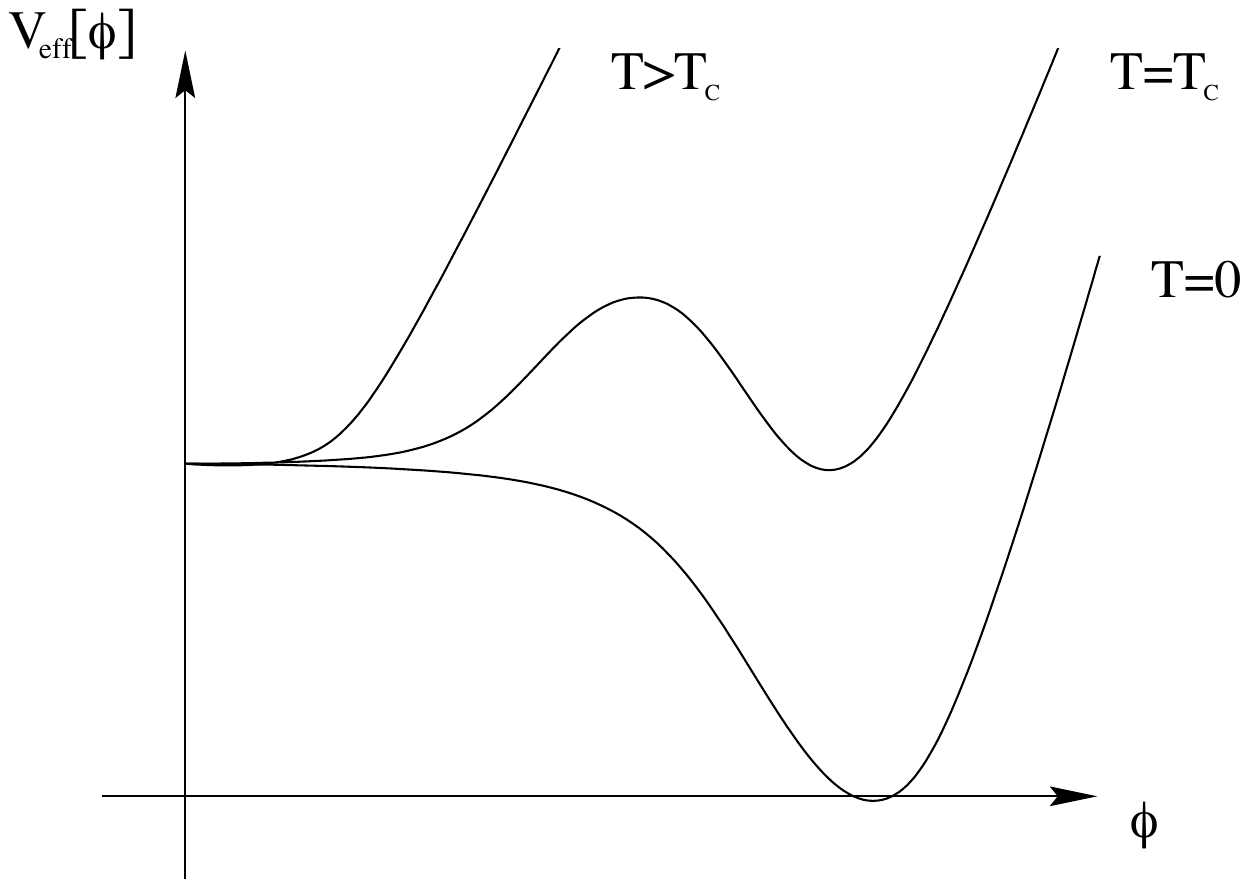}
\caption{First order phase transition. The Higgs vacuum expectation value $v(T)$ varies in a discontinous way at $T=T_c$,
defined as the temperature at which the trivial and non-trivial minima are degenerate.
From Ref.~\cite{Trodden:1998ym}}
\label{fig:Firstorder}
\end{figure}

Finally, non-equilibrium processes occur if the electroweak
phase transition is strongly first order.  In such a case, bubbles of the real vacuum develop in the false, symmetry preserving
vacuum. If a given baryon number is generated at the electroweak phase transition at the critical temperature~\cite{Cohen:1993nk}, after considering
the proper Hubble expansion and the rate of baryon number violation~\cite{Shaposhnikov:1987tw},
$\frac{n_B}{n_\gamma} \sim \frac{n_B}{n_\gamma}(T_c)  \exp\left( -\frac{\Gamma_{\rm Sph}}{H}(T_c)\right)$ ,
where $H$ is the Hubble expansion rate, $H \sim g_{*}^{1/2} T^2/M_{\rm Pl}$, and $g_*$ are the number of relativistic degrees of freedom.  For a transition temperature of the order of 100~GeV,
this can only be fulfilled if
\begin{equation}
\frac{v(T_c)}{T_c} \simgt 1
\end{equation}
implying a strongly first order electroweak phase transition (SFOEPT), which   is schematically represented in Fig.~\ref{fig:Firstorder}

\section{Conditions for a SFOEPT}

We will perform first an
analysis of the Higgs potential at the one-loop level. In such a case,
\begin{equation}
V(h,T) = V_{\rm tree}(h,0) + \Delta V(h,0)+\Delta V(h,T).
\end{equation}
Here $\Delta V(h,0)$ describes the one loop corrections at zero temperature, which include the renormalization of the masses and couplings, and hence depends on the renormalization scale $Q$.  The finite temperature corrections $\Delta V(h,T)$ are finite
and  differ for the case of bosons and fermions, due to their different statistics, and are given by~\cite{Dolan:1973qd} \\
\begin{equation}
\Delta V_{B,F}(h,T) = \pm g_{B,F} T^4 \ I_{B,F}(m_{B.F}(h)/T) \ ,
\end{equation}
where $g_{B,F}$ is the number of degrees of freedom associated with the bosonic and fermionic fields (for instance, four for a Dirac fermion and
one for a real scalar), $I_{B,F}(x)$ is an integral function and  $m_{B,F}(h)$ is the mass of the particle in the background of the Higgs field.

The integral functions may be expressed in terms of a high temperature expansion, which are valid for $m_{B,F}(h) \simlt T$. Ignoring the field indepent terms, $I_{B,F}$ are given by
\begin{equation}
\frac{m_B(h)^2}{24 T^2} - \frac{(m_B^2(h)^{3/2})}{12 \pi T^3} - \frac{m_B(h)^4}{64 \pi^2 T^4} \ln\left[\frac{m_B(h)^2}{T^2}\right]  +  ... ,
\end{equation}
and
\begin{equation}
 \frac{m_F(h)^2}{48 T^2} + \frac{m_F(h)^4}{64 \pi^2 T^4} \ln\left[\frac{m_F(h)^2}{T^2} \right]  + ... \ ,
\end{equation}
respectively.  Observe the appearance of a non-analytic dependence on $(m_B^2(h))^{3/2}$.  In the case
of boson fields with a linear dispersion relation on the Higgs field, this leads to a cubic term on $h$ in the potential that is 
important for the generation of a first order phase transition.

The logarithmic term is of the same form of the logarithmic corrections that appear in the renormalization of the Higgs potential  in the $\overline{MS}$ or $\overline{DR}$ schemes~\cite{Coleman:1973jx}
\begin{equation}
 \pm \frac{g_{B,F} \ m_{B,F}^4(h)}{64 \pi^2} \left( \ln\left[\frac{m_{B,F}^2(h)}{Q^2} \right]+ c \right)
\end{equation}
where $c$ is a constant. This implies that, at one-loop, the logarithmic dependence on $m(h)^2$ cancels at high temperatures.
 
 The Higgs and Goldstone masses become negative at low temperatures, leading to complex contributions to the potential.  
 This can be fixed by a
 Daisy resummation of diagrams~\cite{Dolan:1973qd}, which depends on the thermal mass $m_B^2(\phi,T)$ 
 \begin{eqnarray}
  m_B^2(h) + \Pi_B(T) = m_B^2(h) + a_B T^2 ,
 \end{eqnarray}
 where for any boson field  $a_B$ controls the temperature correction to its mass.  
 After resummation, the bosonic contribution to the Higgs potential are corrected by
 \begin{equation}
  -  T \sum_B g_B  \frac{m_B(h,T)^3-m_B(h,0)^3}{12 \pi}  
 \end{equation}
 leading to a real potential at $T > T_c$. 
 A systematic way of including higher loop effects was presented in Ref.~\cite{Arnold:1992rz}.
 An alternative resummation method, proposed by Parwani~\cite{Parwani:1991gq} is to replace $m_B^2(h)$ by $m_B^2(h,T)$ in
 all terms of the high temperature effective potential.  Away from the high temperature regime one should proceed with care. A   self consistent resummation, in which the Boltzmann suppression factors are included in a natural way has been proposed to deal with this problem (see, for example Refs.~\cite{Curtin:2016urg}). 
  
 Considering the high temperature corrections, assuming that all relevant fields have a linear dispersion relation in $h$ and absorbing the renormalization effects in to a redefinition of the parameters of the model, the Higgs potential then will have the form
 \begin{equation}
 V(h,T) = \frac{(m_H^2 + a_H T^2)}{2} h^2 - E T h^3 +  \frac{\lambda(T)}{8} h^4,
 \label{eq:VhT}
 \end{equation}
 where $E = \sum_B \frac{g_B y_B^3}{12 \pi}$  the sum is performed on all  bosons of  mass $m_B^2 = y_B^2 h^2$, and we have redefined the potential in order to have zero value at $h=0$.  In the Standard Model, the most relevant fields are given by the transverse modes of the gauge bosons~\cite{Carrington:1991hz,Dine:1992vs}.  The contribution to $a_H$, on the contrary, comes from
 all boson and fermion fields.  
 
 As shown in Fig.~\ref{fig:Firstorder},  we shall define the critical
 temperature $T_c$ as the temperature at which this new minimum is degenerate with the trivial one at $h=0$. It is straightfoward to show that this leads to
 \begin{equation}
 \frac{v(T_c)}{T_c} =  \frac{ 4 E } {\lambda(T_c)} .
 \end{equation}
 Considering the mass of the $W$ and $Z$ gauge bosons given by $m_W^2(h) = g_2^2 h^2/4$ and $m_Z^2(h) = (g_2^2 + g_1^2) h^2/4$, respectively,
 and $g_V = 2$ for the transverse modes, we get
 \begin{equation}
 \frac{v(T_c)}{T_c} = \frac{ 2 g_2^3 + (g_1^2+g_2^2)^{3/2}}{12 \pi \lambda(T_c)} 
 \end{equation}
 The requirement that $v(T_c)/T_c \simgt 1$ can only be fulfilled for $\lambda(T_c) \simlt 0.03$, which since $T_c$ is of the order of the weak scale, implies Higgs masses of the order or lower than 40~GeV, which is in conflict with the SM Higgs mass value.   Moreover, a lattice gauge theory analysis shows that the line of first order phase transitions ends at Higgs masses of the order of 70 GeV and for $m_h = 125$ GeV it becomes a crossover~\cite{Kajantie:1995kf,Laine:2012jy}.
 

 \section{New Physics }
 
 The corrections to the Higgs potential  may be induced, for instance, by
 \begin{itemize}
 \item An enhanced cubic term $E$ in $h$,
 \item  A barrier that persists at zero temperature, 
\item Fields that couple strongly to the Higgs (see, for example, Ref.~\cite{Carena:2004ha}). 
\end{itemize}

 The first option can only occur via particles that are present in the plasma, and therefore have masses of the order of the weak scale.   A particular example, that has been studied in much detail was the case of top squarks, which couple with couplings of order one to the Higgs and come in three colors and hence induce additional corrections to the Higgs potential~\cite{Carena:1996wj,Delepine:1996vn,deCarlos:1997ru,Carena:1997ki,Carena:2008vj,Laine:1998qk}.
However, after resummation a large cubic term can only be obtained if  the field independent mass of the stop is negative, implying a very
 light stop. Such a light colored particle  leads to strong contributions to the gluon fusion 
 production of the Higgs boson that have not been observed experimentally~\cite{Cohen:2011ap,Curtin:2012aa}.  Beyond this example, one can postulate other particles to give a finite temperature cubic
 correction, but in general strong couplings and high multiplicities are demanded, that lead to strong constraints on these models~\cite{Katz:2014bha}. 
 In the following, we shall concentrate on the second possibility, that leads to non-trivial modifications of the Higgs properties.  



\subsection{Singlet extension of the SM}

\begin{figure}
\centering
\includegraphics[width= 6.5 cm, clip]{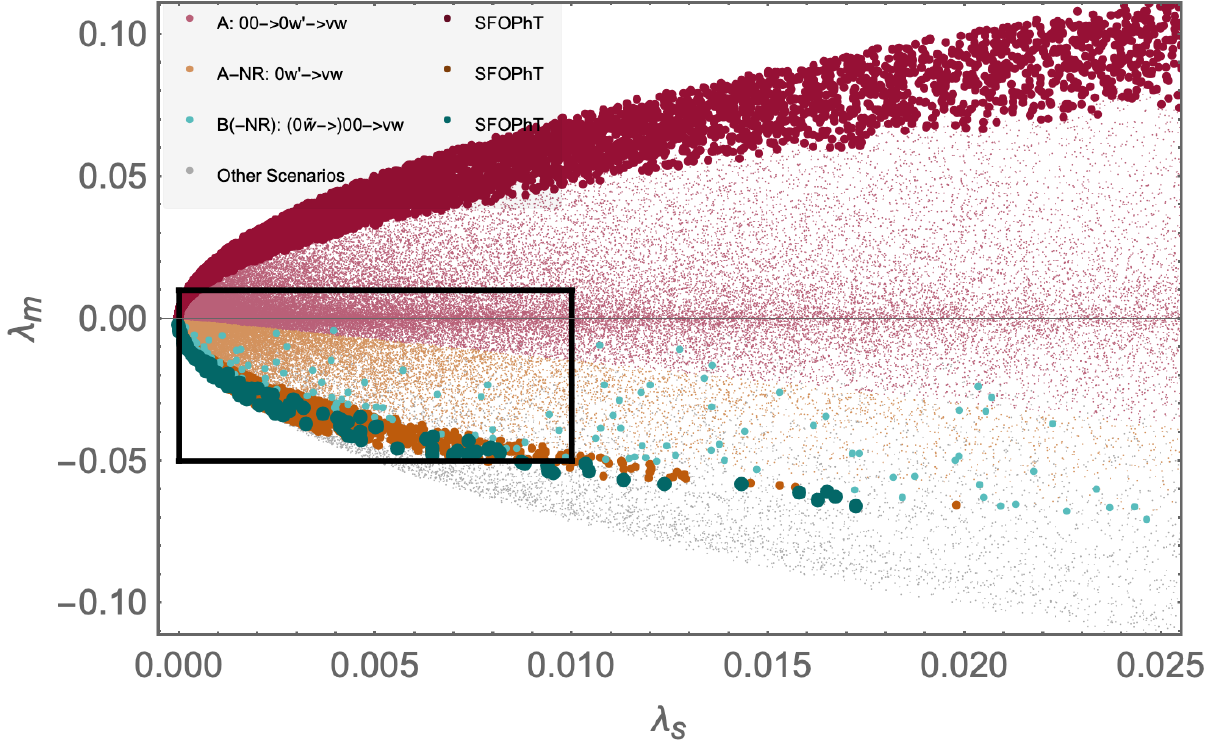}
\caption{Parameters consistent with different transition paterns, where $\lambda_m \equiv \lambda_{sh}$. From
Ref.~\cite{Carena:2019une}}
\label{fig:vclambdah}
\end{figure}
The simplest extension of the SM is just to add a real singlet field. Even this simple extension allows for several new dimensionless and dimensional couplings. A simplification occurs in the case of a discrete $Z_2$ symmetry leading to the addition of the following potential terms,
\begin{equation}
V_{s} = \mu_s^2 s^2 + \frac{\lambda_{sh}}{2}  s^2 h^2 + \frac{\lambda_s}{4} s^4, 
\end{equation}
while the $Z_2$ breaking terms includes linear and cubic terms in the singlet field. An interesting case is the one in which the $Z_2$ symmetry is broken spontaneously. In the simplest approximation in which only the quadratic dependence on the temperature is considered, one obtains that the transition may occur from a minimum with zero doublet and non-zero singlet vacuum expectation value to one in which the doublet expectation value is non-vanishing, which develops at lower temperatures. The critical temperature is then defined when these two minima become degenerate.  It is straightforward to show that~\cite{Carena:2019une}
\begin{eqnarray}
\frac{v_c}{T_c} & = &  \frac{4 E}{ \tilde{\lambda}}= \frac{4E}{\lambda_{\rm SM}} \left[ 1 + \sin^2\theta \left( \frac{m_h^2}{m_s^2} - 1 \right) \right] 
\nonumber\\
  \tilde{\lambda} & =  &\lambda - \frac{\lambda_{sh}^2}{2 \lambda_s} 
\end{eqnarray}
where $\theta$ is the scalar mixing angle.  A strongly first order phase transition may be obtained for a proper choice of $\lambda_{sh}$ and $\lambda_s$.  Such a transition, demands a light singlet and a sizable mixing, which lead to a modification of the couplings of the Higgs to the SM fermions by a factor $\cos\theta$. Large values of $\theta$ are hence
restricted due to precision electroweak and Higgs measurements~\cite{Robens:2015gla}. This opens  the possibility of the SM-like Higgs to decay into two light singlets~\cite{Carena:2019une}.  On the other hand, as it is shown in Fig.~\ref{fig:vclambdah}  in which the full one-loop finite $T$ corrections are included, in general small values of $\lambda_{sh} \simlt {\cal{O}}(0.1)$ and values of $\lambda_s$ which are even
smaller are required in this scenario.

\begin{figure}
\centering
\includegraphics[width=6cm,clip]{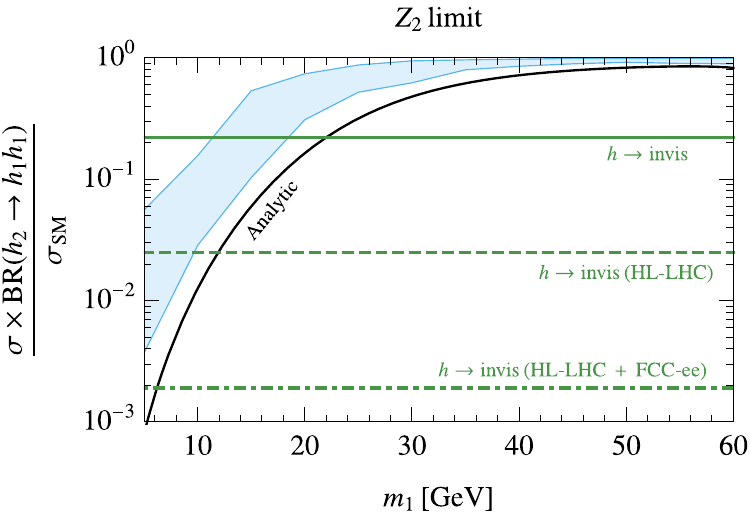}
\caption{Branching ratio for the exotic decay of the SM Higgs going to two light singlets. From Ref.~\cite{Kozaczuk:2019pet}}
\label{fig:BR_Z2.pdf}
\end{figure}

\begin{figure}
\centering
\includegraphics[width=6.5cm,clip]{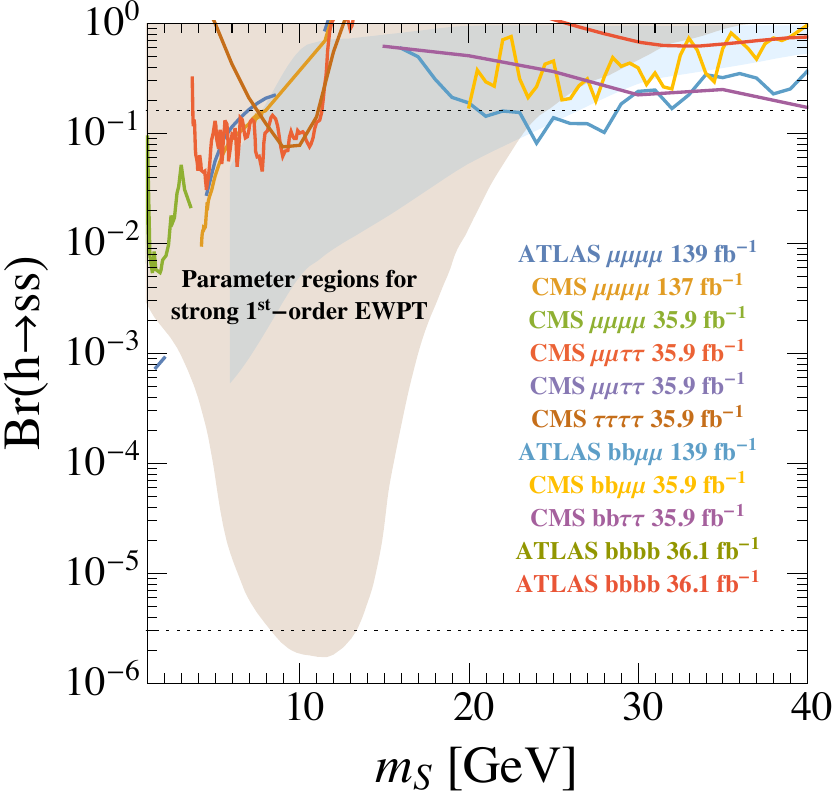}
\caption{Region of parameters for a SFOEPT (shaded region) and the corresponding branching ratio of the decay of the SM-like Higgs into two singlets. Current experimental bounds are given. From Ref.~\cite{Carena:2022yvx}.}
\label{fig:HtoSS}
\end{figure}

The $Z_2$ symmetry may remain preserved at zero temperature via a phase transition leading to 
a breakdown of the $Z_2$ symmetry at high temperatures,  recovered at zero temperature. Certain conditions 
on the couplings must be fulfilled and, in particular, the coupling $\lambda_{sh}$ should be sizable. The results are then less dependent on $E$ but strongly dependent on the structure of the tree-level potential. In the case of a light singlet this may again lead to exotic
decays of the SM-like Higgs. The decay rate is given by~\cite{Kozaczuk:2019pet}
\begin{equation}
\Gamma(h \to s s) \simeq \frac{\lambda_{hs}^2 v^2}{8 \pi^2 m_h} \sqrt{ 1 - \frac{4 m_s^2}{m_h^2}}
\end{equation}
Due to the vanishing scalar mixing, 
the singlet states do not couple to the SM particles and hence these decays contribute to invisible Higgs decays, which are
currently being looked for at the LHC. As it is shown in Fig.~\ref{fig:BR_Z2.pdf}, due to the correlation of the decay branching ratio with the couplings governing
the singlet mass, only light singlets, with mass smaller than about 20~GeV are allowed in this case.
The prediction for and current bounds on the branching ratio of the decay of a SM-like Higgs boson into two singlets, for a more general
case of a SFOEPT in singlet extensions of the SM is depicted in Fig.~\ref{fig:HtoSS}.

One can also consider singlet extensions, including other symmetry breaking patterns and phenomenological properties, like Dark Matter and we refer the reader to 
Refs.~\cite{Espinosa:1993bs,Choi:1993cv,Profumo:2007wc,Barger:2011vm,Cline:2012hg}, for more
details on these subjects.). 
 
\subsection{Heavy Singlet field}
\label{sec:EWPT}

In this case,  we will not demand a $Z_2 $ symmetry and we shall
integrate out the singlet field. We shall include the potential terms
\begin{equation}
V(h,S) = \frac{m_H^2}{2} h^2 + \frac{\lambda}{8} h^4  + \frac{m_S^2}{2} s^2  + A_{hS} s h^2 + \frac{\lambda_{hS}}{2} h^2 s^2 
\end{equation}
The equation of motion of the field $S$ is, at low momentum, given by their interaction with the 
Higgs field $h$~\cite{Menon:2004wv,Carena:2011jy,Huang:2015tdv}
\begin{equation}
S = - \frac{A_{hS} h^2}{m_S^2 + \lambda_{Sh} h^2}
\end{equation}
Assuming that $m_S$ is large compared to the weak scale, one can therefore obtain an effective potential by replacing $s$ in the
previous expression
\begin{equation}
V_{\rm eff}(h) =  \frac{m^2}{2} h^2 + \frac{\lambda}{8} h^4 - \frac{ (A_{hS} h^2)^2}{2(m_{S}^2 + \lambda_{hS} h^2)}
\end{equation}

Dxpand $V_{\rm eff}(h)$ in powers of $1/m_S^2$, 
\begin{equation}
\frac{m^2}{2} h^2 + \left( \frac{\lambda}{8}  - \frac{A_{hS}^2}{2 m_S^2} \right) h^4 +  \frac{A_{hS}^2 \lambda_{hS}}{2 m_S^4} h^6 +... \ ,
\end{equation}
the result is then an effective potential in which the quartic term is modified and may become negative, and there are non-renormalizable
corrections which are parametrized by $m_S^2$. and the trilinear and quartic couplings $A_{hS}$ and $\lambda_{hS}$. If one defines the correction to be proportional to $c_6 h^6/( 8 \Lambda^2)$, however, this cutoff has a non-trivial dependence on these couplings and cannot be identified with the mass scales $m_S$. If we go to higher orders in $1/m_S^2$, one will obtain higher powers of $h$ in the potential. We will provide a more general expression in terms of an effective theory in the next section. 

\subsubsection{Higgs Potential at higher orders in $h^2$}
\label{sec:phi6}

The simplest extension would be given by
\begin{align}
V(\phi,T)&=\frac{m^2+a_0T^2}{2}h^2+\frac{\lambda}{8} h^4+\frac{c_6}{8\Lambda^2}h^6\label{eqn:temp6}
\end{align}
We do not include a cubic $E$ term, assuming that its effects are small compared to the dominant temperature effects.
This case has been studied in the literature in various contexts~\cite{Barger:2011vm, Chung:2012vg,Noble:2007kk,Katz:2014bha,Curtin:2014jma,He:2015spf,Grojean:2004xa,Huang:2015tdv,Huang:2016cjm}. 
It is straighforward to show that the Higgs mass and the trilinear Higgs coupling are given by
\begin{align}
m_h^2 & = \lambda v^2 + 3 \frac{c_6 v^4}{\Lambda^2} \nonumber\\
\lambda_3 &=\frac{3m_h^2}{v}\left(1+\frac{2c_6v^4}{m_h^2\Lambda^2}\right)\label{eqn:enh6}
\end{align}
Therefore, there is a relevant modification of the trilinear Higgs coupling, proportional to $c_6$,  that is one of the most relevant signatures of this scenario. 
We require $c_6>0$ for the stability of the potential. 
The requirement of a minimum of the potential at $h=v(T_c)=v_c$ degenerate with the extreme at $h= 0$ at $T=T_c$ leads to 
\begin{align}
v_c^2=-\frac{\lambda \Lambda^2}{2 c_6}.\label{eqn:root2}
\end{align} 
what implies $\lambda < 0$. 
%
Using Eq.~(\ref{eqn:enh6}) and Eq.~(\ref{eqn:root2}), and the minimization relations, one obtains
\begin{align}
\frac{c_6}{\Lambda^2}=\frac{m_h^2}{3v^2\left(v^2-\frac{2}{3}v_c^2\right)}\label{eqn:kappa} ,
\end{align}
 %
\begin{align}
T_c^2=\frac{3 c_6}{4 \Lambda^2 a_0}&\left(v^2-v_c^2\right)\left(v^2-\frac{v_c^2}{3}\right).\label{eqn:critical1}
\end{align}
Demanding both $c_6$ and $T_c^2$ to be positive, we get that  trilinear coupling in the case of a first order phase transition is bounded by
\begin{eqnarray}
\label{eq:enhancement}
\frac{2}{3}   \leq  &  \delta  &  \leq  2 , 
\end{eqnarray}
where $\delta =  \frac{\lambda_3 - \lambda_3^{\rm SM}}{\lambda_3^{\rm SM}} = \kappa_\lambda -1$.

\begin{figure}[ht]
\centering
\includegraphics[width=6.5cm,clip]{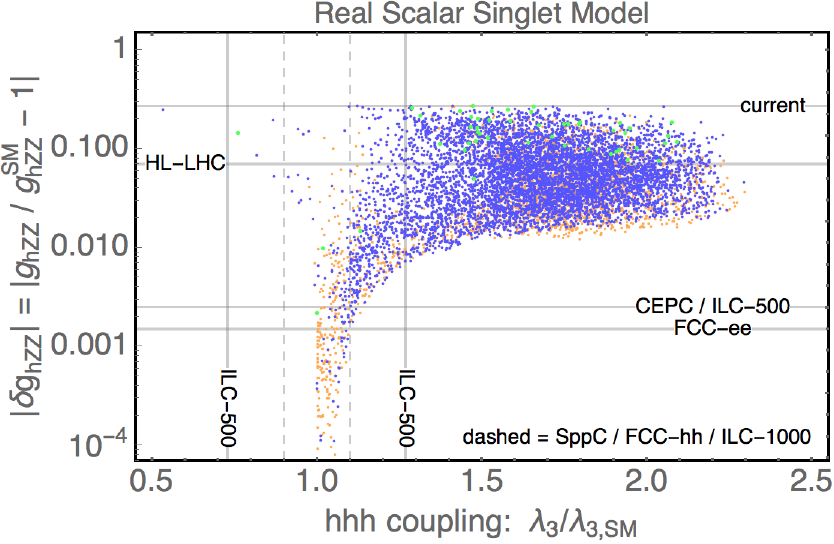}
\caption{
Modifications of $\lambda_3$ and of $g_{hZZ}$ for the case of a (very) strongly first order phase transition
are shown by the (green) blue points. From Ref.~\cite{Huang:2016cjm}}
\label{fig:lambda3SFOEPT}
\end{figure}

Moreover, for $c_6=1$, we obtain a bound on the effective cutoff $\Lambda$, namely
\begin{eqnarray}
\frac{v^2}{m_h}  < &  \Lambda &  < \frac{\sqrt{3}v^2}{m_h}  ,
\label{eq:enhancement2}
\end{eqnarray}
which correspond to upper and lower bounds on $\Lambda$ of approximately 500 GeV and 850 GeV respectively, and as shown in Eq.~(\ref{eqn:enh6}) larger enhancement $\delta$ is obtained for the smaller values of the cutoff. The phase transition becomes stronger first order for smaller values of the cutoff
and becomes a weakly first order one for values of $\Lambda$ close to the upper 
bound in Eq.~(\ref{eq:enhancement2}). 
\begin{figure}[ht]
\centering
\includegraphics[width = 6.5cm, clip]{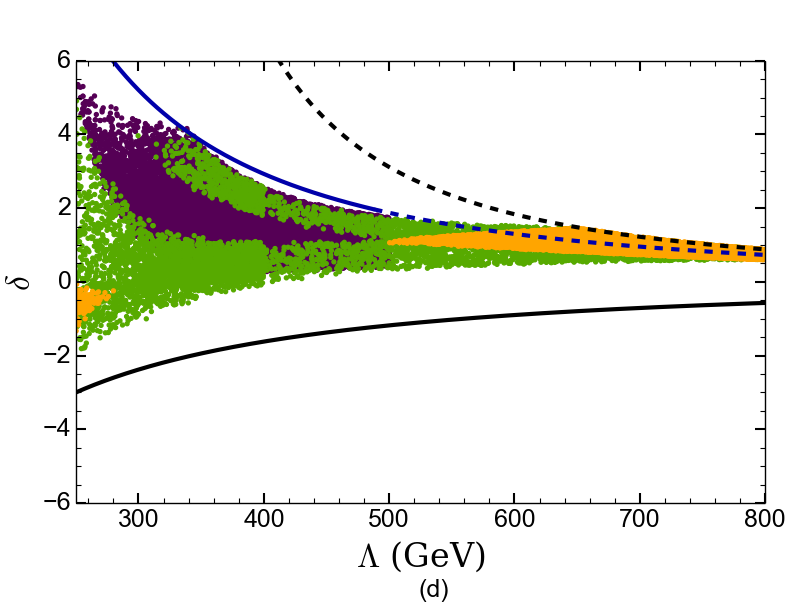}
\caption{Triple Higgs coupling correction $\delta$ as a function of the cutoff $\Lambda$ for the $h^{10}$ case. The dashed dark blue shows the values consistent with a first order electroweak phase transition  FOEPT for the $h^6$ potential extension, for $c_6 =1$. The different colors correspond to the different hierarchies of the effective potential $h^{2n}$ coefficients in the case of a FOEPT.  For other details see Ref.~\cite{Huang:2015tdv}.}
\label{fig:convergence}
\end{figure}
In Fig~\ref{fig:convergence}, we show the possible triple Higgs coupling enhancement factor $\delta$ as a function of the cutoff $\Lambda$ for different extensions of the SM effective potential. The particular case of the potential of order $h^6$ is represented by the blue curve. The conditions to obtain a FOEPT up to order $h^{10}$ are shown in this Figure.  The parameter space consistent with a SFOEPT are shown in Fig.~\ref{fig:lambda3SFOEPT}.


\subsection{Experimental tests of the Trilinear Higgs coupling}

The determination of the trilinear Higgs coupling $\lambda_3$ will be the first probe of the Higgs potential beyond the ones provided by the Higgs expectation
value and the Higgs mass, with determine the position of its minimum and the associated curvature. Any deviation
of the Higgs potential with respect to the SM prediction, as demanded by electroweak baryogenesis, will lead to a modification of $\lambda_3$ and therefore the determination of the trilinear Higgs coupling will be an excellent probe of this scenario.

\begin{figure}[tbh]{
\centering
\includegraphics[width = 6.5cm, clip]{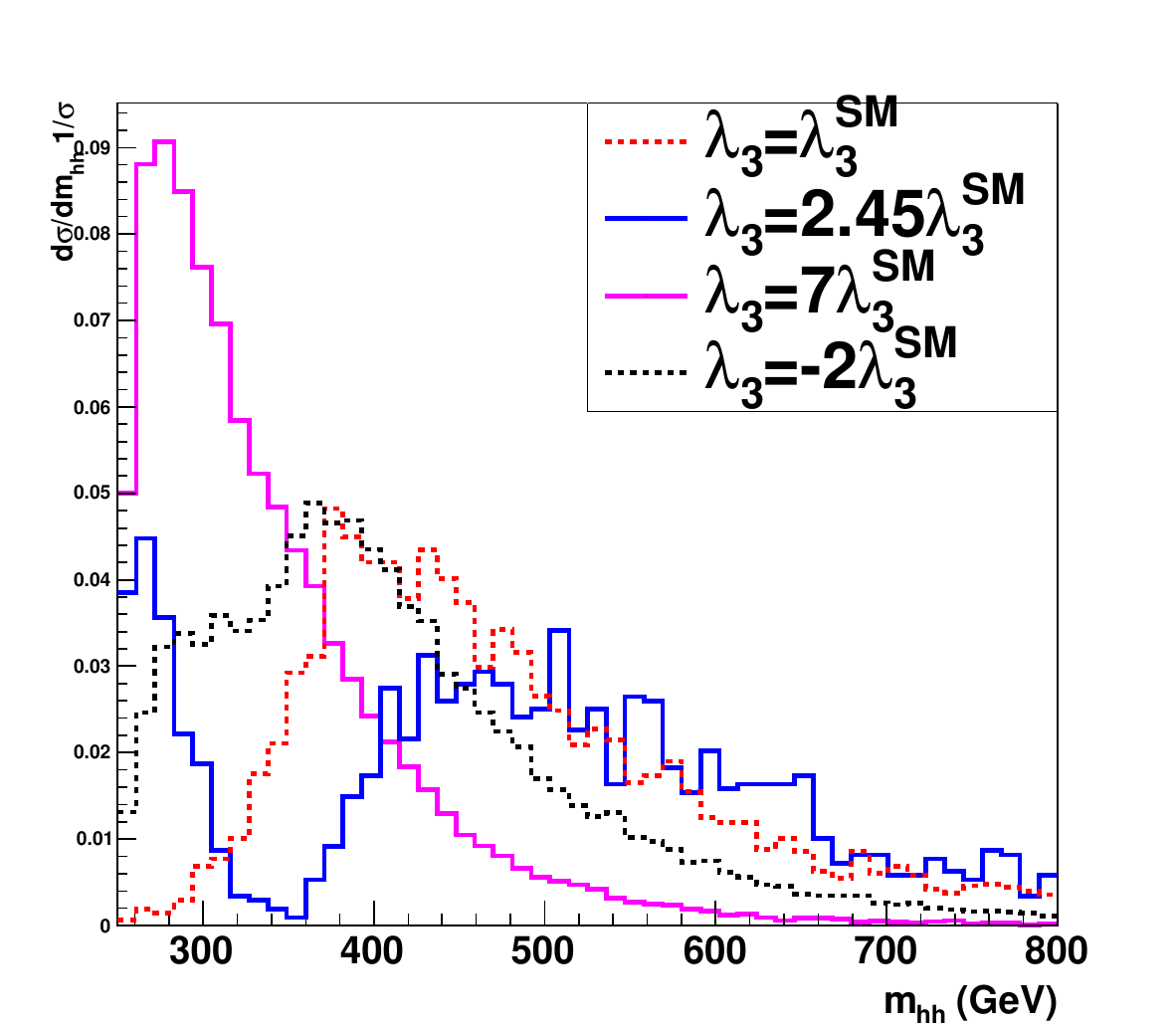}
\caption{Normalized $m_{hh}$ distributions for 
different values of $\lambda_3$ . The cancellation between the box and triangle diagram is exact at $\lambda_3$ = 2.45$\lambda_3^{SM}$ at 2$m_t$ threshold, what explains the dip. Note that the distribution shifts to smaller values as $|\lambda_3|$ increases. From Ref.~\cite{Huang:2017nnw}.} 
\label{fig:mhh}}
\end{figure}

The Higgs trilinear coupling $\lambda_3$ can be probed by the double Higgs production at the LHC. At the leading order (LO), there are two diagrams contributing to the process. The triangle diagram, which is sensitive to $\lambda_3$ and the box diagram. The two diagrams interfere with each other destructively. The QCD NNLO differential cross sections are known~\cite{Glover:1987nx,Dicus:1987ic,Plehn:1996wb,Dawson:1998py,Kniehl:1995tn,KFactor,deFlorian:2013jea,Grigo:2014jma,deFlorian:2016uhr,deFlorian:2017qfk,Davies:2018qvx}. 
For the Higgs decays, one can consider $\gamma\gamma$, $W^+W^-$, $\tau^+\tau^-$ and $b\bar{b}$ modes, which are measured in the single Higgs production at the LHC.   
As shown in Fig.~\ref{fig:mhh}., the $m_{hh}$ predicted range depends strongly on the coupling $\lambda_3$ , and therefore the precise experimental window must be fixed judiciously in order to increase the search sensitivity~\cite{Baglio:2012np, Barger:2013jfa, Yao:2013ika, Huang:2015tdv}.  

\begin{figure}
\centering
\includegraphics[width= 6.5  cm, clip]{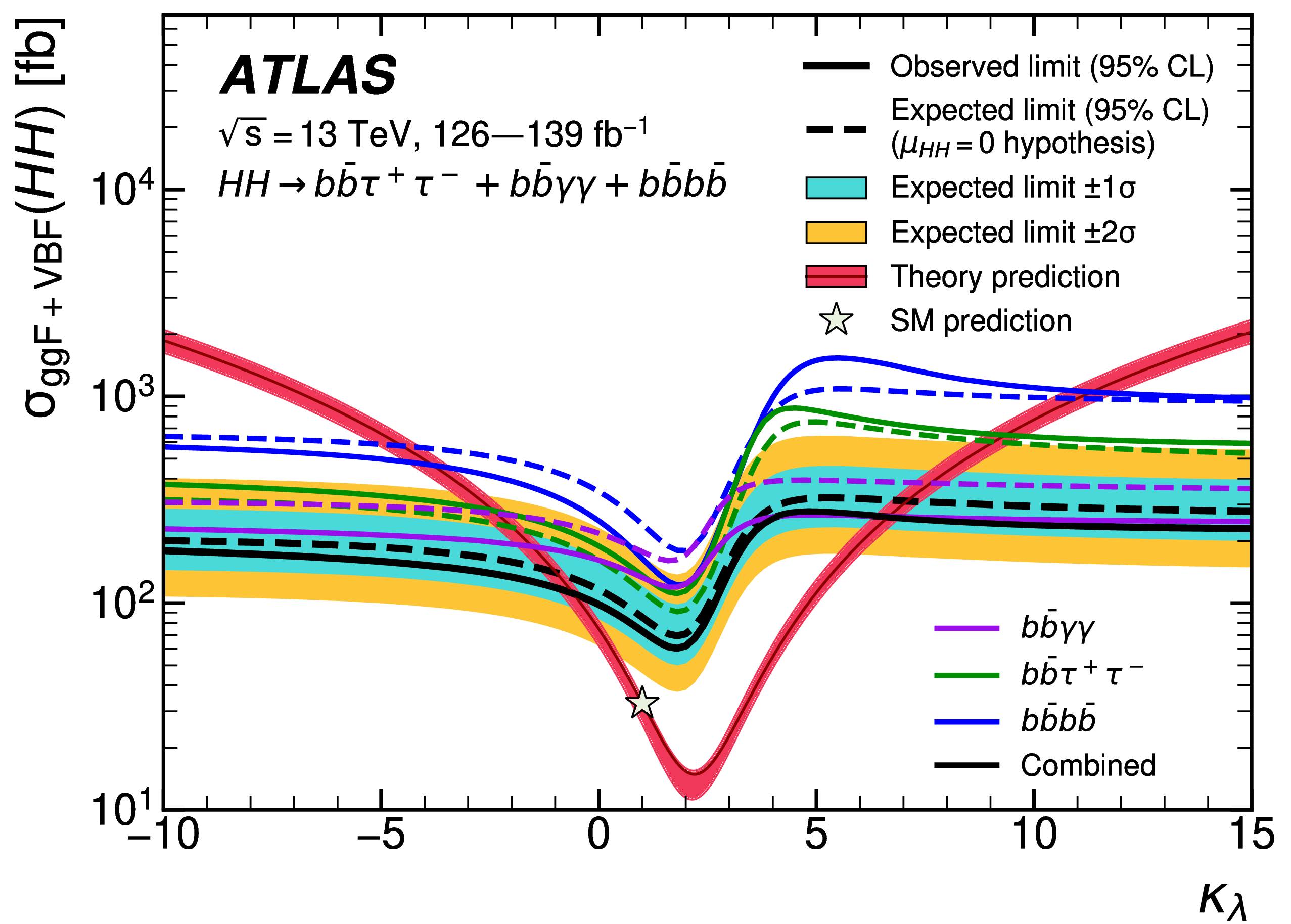}
\caption{Current sensitivity in the determination of the triple Higgs coupling, assuming all other couplings to be fixed to the SM value. From Refs.~\cite{ATLAS:2022jtk}}
\label{fig:tripleHiggssens}
\end{figure}

The current sensitivity of di-Higgs production at the ATLAS experiment in different channels is shown in Fig.~\ref{fig:tripleHiggssens}.  Due to the sensitivity of the di-Higgs production on the top Yukawa coupling (see, for instance, Ref.~\cite{Huang:2017nnw}), the determination of 
$\kappa_\lambda$ has to be made by a fit to the single and di-Higgs production, but the allowed range of $\kappa_\lambda$ is similar to the one obtained in the $\kappa_t = y_t/y_t^{\rm SM} = 1$ case~\cite{ATLAS:2022jtk}.
%

The prospects for the high luminosity LHC run with 3000~fb$^{-1}$ are excellent.  Assuming an improvement of a factor 2 on both theoretical and systematic errors, something not unexpected,
the sensitivity of the experiments for non-resonant Higgs production will be of about half of the SM cross section, 
implying a 3.4~$\sigma$ evidence for double Higgs production in the case of a SM cross section~\cite{ATLAS:2022faz},
and without taking into account the necessary combination of ATLAS and CMS analyses. Therefore, in combination with the single Higgs production cross section, which will strongly constrain the Higgs coupling to top and gauge bosons, the di-Higgs production will become an excellent probe of trilinear Higgs coupling modifications and therefore of electroweak baryogenesis.

Let us say, in addition, that resonant decays of non-standard Higgs bosons into pairs of SM-like Higgs bosons are also an important probe of baryogensis scenarios. We refer to Refs.~\cite{No:2013wsa,Chen:2014ask} for a further discussion of this subject.  Finally, let us also stress that non-resonant di-Higgs production may be also modified by the presence of extra color particles in the loop diagram, like in the cases studied in Refs.~\cite{Huang:2017nnw, Moretti:2023dlx, DaRold:2021pgn}.

\subsection{Extensions with extra Higgs Doublets}

Two Higgs doublet models have also been studied in detail and the parameter space consistent with a
strongly first order phase transition has been determined by both analytical and numerical methods, see for example~\cite{Dorsch:2013wja,Basler:2016obg,Basler:2021kgq}.  A  relevant parameter in these models is the
ratio of the two Higgs vacuum expectation value, $\tan\beta$.
Deviations of the SM-like Higgs boson couplings with respect to the SM value in the CP conserving case are controlled by the misalignment
of the direction of the Higgs expectation value, controlled by $\beta$, with the mixing in the CP-even sector, $\alpha$.
Values of $\cos(\beta-\alpha) \sim 0$ lead to the presence of a light Higgs  with SM-like couplings~\cite{Gunion:2002zf,Carena:2013ooa}, and this condition tends to be fixed in most simulations. In the CP-violating case, additional mixing angles appear and the condition of 
alignment can be easily generalized to this case. 

For non-standard Higgs bosons with masses which are larger than the weak scale, preferred from
the point of view of LHC searches, large quartic couplings are necessary in order to get a SFOEPT.
These large couplings tend to induce a splitting between the CP-even and CP-odd Higgs bosons,
which are assumed to be approximately good eigenstates, leading to the possibility of the
decay of $A \to H Z$, which can be searched for at the LHC~\cite{Dorsch:2014qja,Biekotter:2023eil}.    Recently, there has been also an intriguing
claim of the possibility of more complicated phase transition patterns in the two Higgs doublet models, including
the possibility of symmetry non-restoration, which would be very interesting and must be elucidated by further work~\cite{Biekotter:2021ysx}.

The addition of doublets and singlets has also been studied, for instance in the context of the Next To Minimal extension of the SM. This leads to a very complicated pattern of phases, which includes the possibility of a SFOEPT for particular values
of the trilinear couplings $A_{\lambda}$ of the singlet-Higgs fields $A_{\lambda} S H_u H_d$,  even in the case of small quartic couplings~\cite{Pietroni:1992in,Davies:1996qn,Huber:2006wf,Huang:2014ifa,Baum:2020vfl}. The abundances of Higgs field directions imply the existence of several minima and transitions from one to the other may be obstructed by large barriers which demand a careful analysis of the transition rates in order to determine the possibility of baryogenesis~\cite{Baum:2020vfl}. Appropriate conditions may be found in regions of parameter space consistent with Dark Matter and a SM-like Higgs boson.  Exotic decays of the heavy Higgs bosons into light singlets and SM Higgs and gauge boson states are expected in this case~\cite{Carena:2015moc}.

Further interesting scenarios include the possibility of inert doublets and composite Higgs models. We refer to
Refs.~\cite{Espinosa:2011eu,Blinov:2015sna,Bruggisser:2018mrt} for further details on these subjects. 

\section{CP-Violation}

\subsection{Experimental Constraints}

As stressed above, new CP violating sources are necessary in order to generate the observed baryon asymmetry.
These CP-violating phases tend to lead to contributions to the SM fermion electric dipole moments. These contributions
depend on the interactions of these particles with the SM particles, but parametrically, for the electric dipole moments, they are governed by~\cite{Morrissey:2012db}
\begin{eqnarray}
d_f & \sim & sin(\theta) \frac{\alpha}{4\pi} \frac{m_f}{M^2}  
\nonumber\\
d_f & \sim & \sin(\theta) \frac{ m_f }{ {\rm MeV} }  \left( \frac{ 1 {\rm TeV} }{M}  \right)^2 \times 10^{-26} e \cdot {\rm cm} ,
\end{eqnarray}
where $\theta$ is a new CP-violating phase, and $M$ are the masses of the new particles. Observe that the constant of proportionality hides the details of the particle interactions. However, the comparison of this prediction with the current experimental bound on the electric dipole moment of the electron~\cite{ACME:2018yjb,Roussy:2022cmp}
\begin{equation}
d_e < 4.1 \times10^{-30} e \cdot cm
\end{equation}
implies a strong bound on the new CP-violating sector. 

The LHC can also put bounds on the CP violating effects, by measuring the Higgs rates, that are affected by
the inclusion of CP violating couplings, or directly by looking by CP violating effects in the production and
decay processes of the SM Higgs (see, for instance, Refs.~\cite{Bahl:2022yrs,Berge:2011ij,MorenoLlacer:2021uos,CMS:2021sdq,ATLAS:2020evk,ATLAS:2022akr}).

\subsection{CP Violating Sources}

\begin{figure}
\centering
\includegraphics[width=6cm,clip]{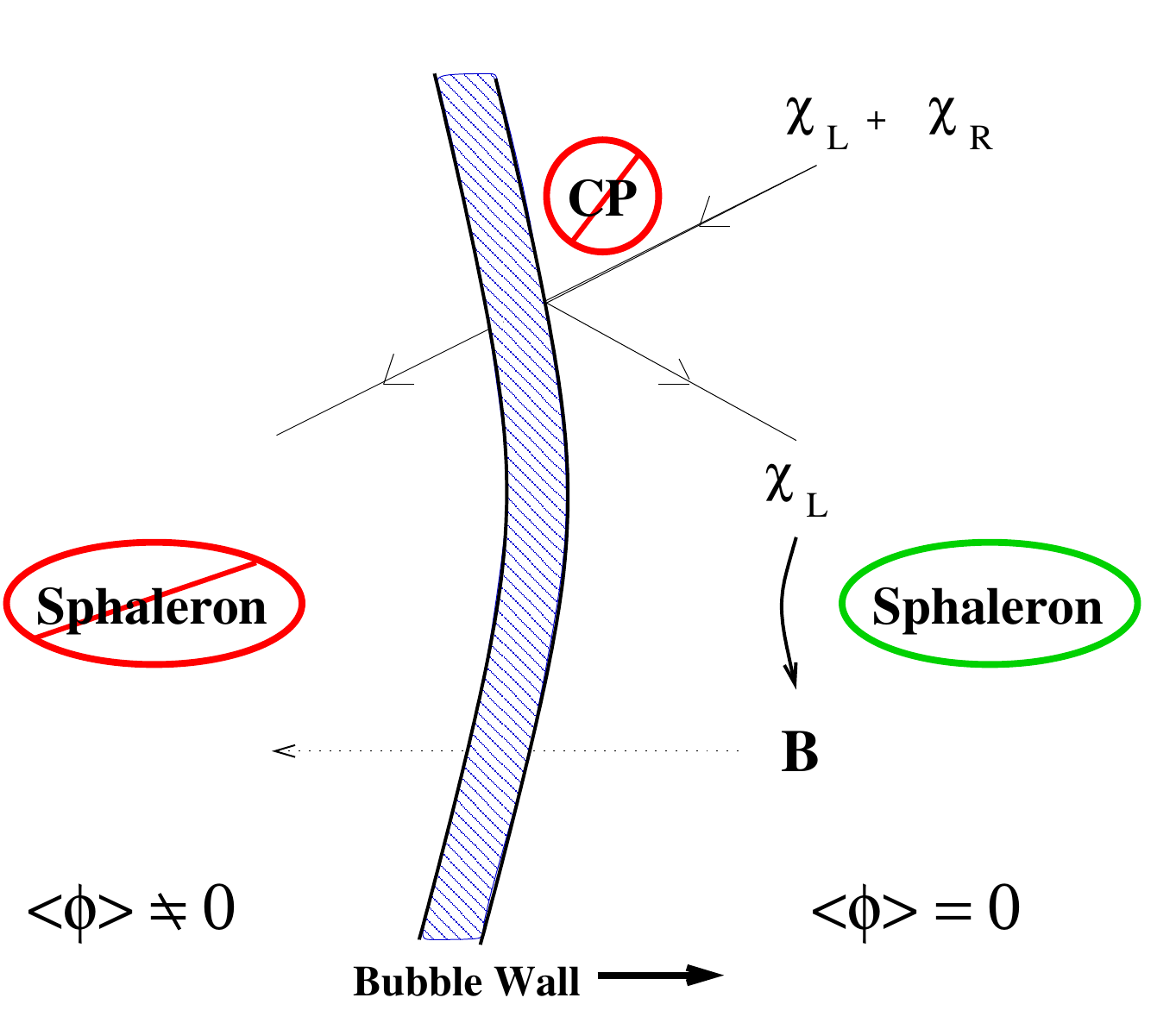}
\caption{Particle transmission at the bubble wall. Inside the bubble, the sphaleron processes are suppressed.
From Ref.~\cite{Morrissey:2012db}}
\label{fig:bubble}
\end{figure}

The way the new CP-violating sources impact the baryon asymmetry is through the generation of particle CP-violating densities at the walls of bubbles of real vacuum state that expand at a certain velocity $v_w$. The particle densities diffuse into the bubbles while the left-handed baryons interact with the sphaleron states that provide the necessary baryon (and lepton) number violating processes.  This is depicted in Fig.~\ref{fig:bubble}. One should therefore solve the system of diffusion equations for the number densities or chemical potentials $\mu_i$
\begin{equation}
D_i \mu_i^{''} - v_w \mu_i + \Gamma_{ij} \mu_j = S_i
\end{equation}
where $D_i$ is the diffusion constant, $\Gamma_{ij}$ is the rate of interactions of the particle $i$ with other particles $j$ ,
the primes indicate derivatives with respect to the direction orthogonal to the bubble wall and $S_i$ are the CP-violating sources (see, for instance, Ref.~\cite{Huet:1995sh}). Although the sphaleron processes in the symmetric phase should be included among the $\Gamma_{ij}$, they tend to be much slower than the other interactions $\Gamma_{\rm Sph} \sim \alpha_w^5 T$ ($\alpha_w$ is the weak coupling constant) and it is convenient to solve first for the CP-violating sources and finally solve for the baryon asymmetry while considering their sphaleron interactions.  The final baryon asymmetry is
then given by
\begin{equation}
n_B  \sim - \frac{\Gamma_{\rm Sph}}{v_w} \int_{-\infty}^0 n_{B_L}(z) \exp(z R/v_w) 
\end{equation}
where $R \propto \Gamma_{\rm Sph}$ is a relaxation coefficient and $n_B$ is the final, constant value in the broken phase.   It is clear from this equations that the bubble wall velocity (and the bubble wall width)
plays a relevant role on these predictions and that $S_i$ must be properly computed in order to predict the correct baryon number. Both calculations are theoretically challenging (see, for instance Refs.~\cite{Huet:1995sh,Riotto:1995hh,Carena:1997gx,Cline:1997vk,Carena:2000id,Cline:2000nw,Kainulainen:2001cn,Carena:2002ss,Balazs:2004bu,Konstandin:2005cd,Huber:2006wf,Postma:2022dbr} and ~\cite{Moore:1995si,Huber:2011aa,Wang:2020zlf} ) and 
 have been done in different approximations that for physics
at the weak scale and phases of order one tend to predict values of the baryon asymmetry of the order of the observed values. Let us stress
that there are large theoretical uncertainties in the computation of these quantities and therefore baryogenesis results are indicative of the region of parameters preferred in a given scenario, but should not be taken as precise predictions of the models under analysis. 

\subsection{Modified Higgs Couplings}

 Since the mechanism of baryogenesis is connected to 
the electroweak phase transition, it is naturally to expect that the Higgs may inherit CP-violating effects, including
CP-violating couplings.  If only the SM particles are in the high temperature plasma at low energies, one can
interpret the baryogenesis as being induced by the difference of the  particle and anti-particle transmission and reflection coefficients at the buble wall, induced by the CP-violating effects in the Higgs sector~\cite{Kobakhidze:2015xlz, Guo:2016ixx,  deVries:2017ncy, DeVries:2018aul,Fuchs:2020uoc}.  

Following Ref.~\cite{Fuchs:2020uoc}, one can define the Lagrangian
\begin{equation}
{\cal{L}} =   y_f \left(1  +   \frac{2}{v^2}  (T_R^f + i T_I^f) H^\dagger H \right)   \bar{f}_L f_R H + h.c. 
\end{equation}
where $T_R^f$ and $T_I^f$ characterize the Higgs fermion coupling modifications. In particular,
\begin{equation}
 \frac{y_f}{y_f^{\rm SM}} = \frac{1}{\sqrt{ (1 + T_R^f)^2 + (T_I^f)^2}}.
\end{equation}
The CP-violating sources depend on the allignment of the fermion mass with respect to its derivative with respect to the coordinate perpendicular to the bubble wall.  The final result is given approximately by~\cite{Fuchs:2020uoc,Fuchs:2020pun}
\begin{equation}
Y_B=8.6\times10^{-11}\times(51 T_I^t - 23 T_I^\tau - 0.44 T_I^b),
\label{eq:YBttaub}
\end{equation}
where $Y_B = n_B/s$  and $s$ is the entropy density, $s \sim 7 n_\gamma$
Hence, it is very
important to determine the bounds on the imaginary component $T_I^f$. These bounds originate mainly from the 
electron dipole moment, which receives a contribution at two loops from the third generation Higgs fermion couplings,
$d_e \approx 1.1 \times 10^{-29} \;e\; \text{cm}\; $ times approximately
\begin{equation}
2200 \left(\frac{y_t}{y_t^{\rm SM}}\right)^2 T_I^t
+ 10 \left(\frac{y_\tau}{y_\tau^{\rm SM}}\right)^2 T_I^\tau
+ 12 \left(\frac{y_b}{y_b^{\rm SM}}\right)^2 T_I^b 
\label{eq: d_e coefficients}
\end{equation}
\begin{figure}
\centering
\includegraphics[width = 6. cm, clip]{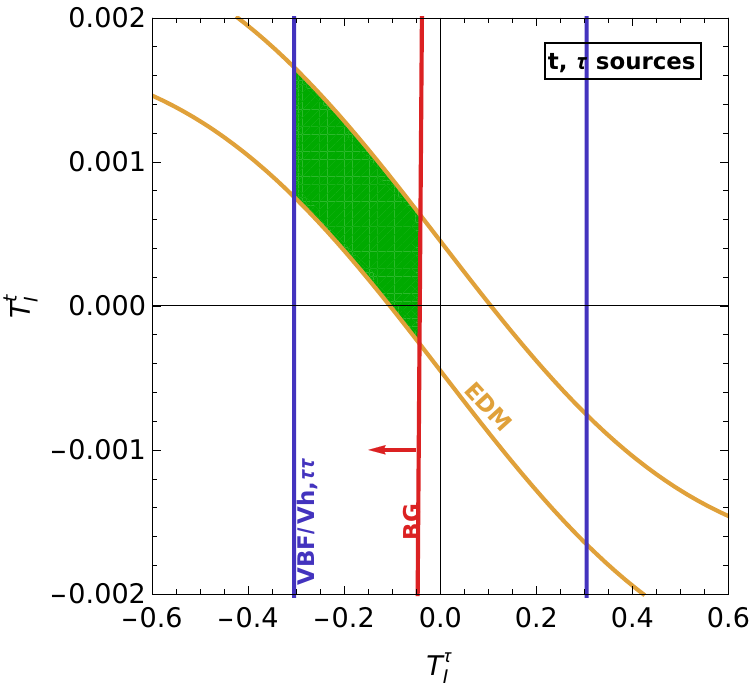}
\caption{Constraints coming from electric dipole moments, Higgs precision measurments and the requirement of Baryogenesis. From Ref.~\cite{Fuchs:2020uoc}.}
\label{fig:BaryoCP}
\end{figure}
It is clear from Eqs.~(\ref{eq:YBttaub}) and (\ref{eq: d_e coefficients}) that the tau coupling modification is the most likely to lead to a non-vanishing baryogenesis contribution. Single top or bottom Yukawa coupling effects can lead to only a fraction of the total baryon number, but can add relevantly to the tau contribution. An example is given in Fig.~\ref{fig:BaryoCP}. 

\begin{figure}
\centering
\includegraphics[width=7.5cm,clip]{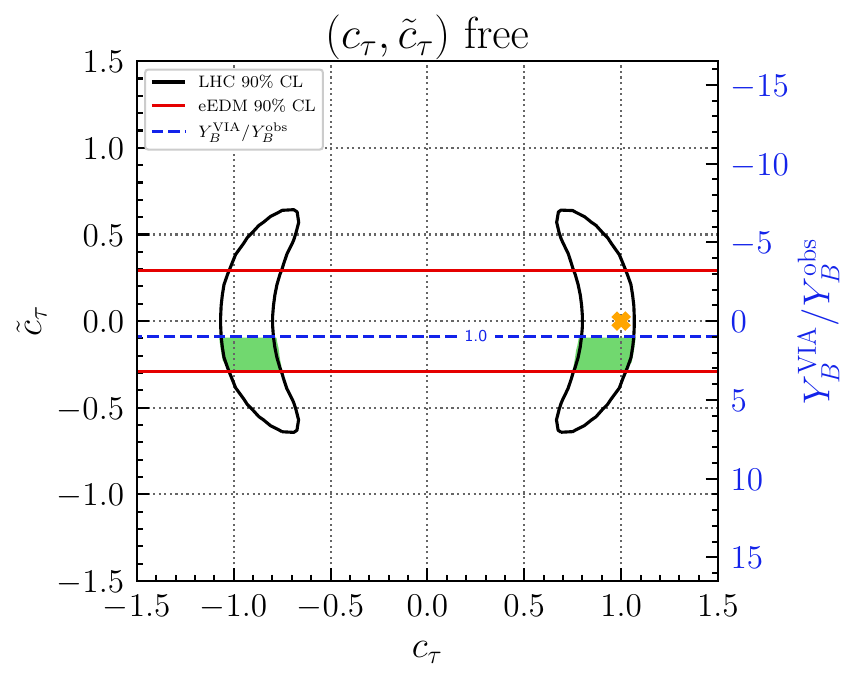}
\caption{Constraints on the tau coupling from baryogenesis, LHC Higgs data and electric dipole moments. Here $c_\tau = T_R^{\tau}$ and $\tilde{c}_\tau = T_I^{\tau}$. From Ref.~\cite{Bahl:2022yrs}.}
\label{fig:CPHFit}
\end{figure}

 As it is shown in Fig.~\ref{fig:CPHFit} these coupling modifications are not yet constrained by the LHC~\cite{Bahl:2022yrs},
 but may be probed in the near future at this collider~\cite{Berge:2011ij,MorenoLlacer:2021uos,CMS:2021sdq,ATLAS:2020evk,ATLAS:2022akr}. 

\subsection{CP Violation beyond the SM Higgs Sector}

Let us stress that the above analysis is only valid if only the third generation SM particles participate in the process of baryogenesis and in the contribution to electric dipole moments.  This is generically not the case and therefore a more complete analysis is in order. In two Higgs doublet models~\cite{Cline:1995dg,Cline:2011mm,Fromme:2006cm,Basler:2021kgq}, for instance, the electric dipole moments receive contributions from both the standard and non-standard Higgs bosons (for a complete analysis in the case of a $Z_2$ symmetry see Ref.~\cite{Altmannshofer:2020shb}).  There are also specific correlations between the different CP-violating couplings that must be taken into account in different kind of two Higgs doublet models and cancellations between different contributions to the electric dipole moments~\cite{Bian:2014zka,Shu:2013uua} may occur, that can lead to the possibility of baryogenesis in these models.  However, as emphasized in Ref.~\cite{Inoue:2014nva} cancellation effects on the electric dipole moment of the electron may not exclude the constraints from the neutron or Hg electric dipole moments and hence the parameter space should be carefully studied to evaluate the possibility of electroweak baryogenesis. \\

In supersymmetric extensions, the CP-violating sources include relevant contributions from the electroweak fermion sector, as has been shown in many relevant works~\cite{Huet:1995sh,Riotto:1995hh,Carena:1997gx,Cline:1997vk,Carena:2000id,Cline:2000nw,Kainulainen:2001cn,Carena:2002ss,Balazs:2004bu,Konstandin:2005cd,Huber:2006wf,Postma:2022dbr}, and this scenario is consistent with the inclusion of a Dark Matter candidate.  The CP-violating currents originate from contributions coming from the relative phases of the gaugino and Higgsino mass parameters and are proportional to derivatives of the Higgs fields, which imply non-standard Higgs bosons at the reach of the LHC. Similarly to the case of two Higgs doublet models, the electric dipole moments lead to strong constraints on these models, unless specific cancellations between different contributions occur~\cite{Brhlik:1998zn,Ibrahim:1999af}.

Let us add in closing that the strong constraints imposed by electric dipole moments have motivated the study of additional baryogenesis scenarios in which CP violation occurs in a Dark sector (see Refs.~\cite{Cline:2017qpe,Carena:2018cjh,Carena:2019xrr}). These scenarios are associated with reduced, but still non-vanishing electric dipole moments that may be probed in future searches.  A related idea is the possibility of breaking CP at temperatures close to the phase transition, enabling the baryogenesis processes, while recovering the CP symmetry at zero temperature, leading to no observable electric dipole moments. This idea has been explored in Refs.~\cite{Huber:2022ndk,Biermann:2022meg},
and the conditions for baryogenesis were  further studied in Ref.~\cite{Huber:2022ndk}. 

\section{Conclusions}
Bayogenesis at the electroweak phase transition remains as an intriguing possibility and demands new physics at the weak scale. This scenario can be probed by precision Higgs physics, exotic Higgs decays as well as double Higgs production. 
Some of these probes, related to the presence of light singlets as well as the modification of the trilinear Higgs couplings, were discussed in this review. Quite generally,
since the new physics inducing these modifications should not decouple, one must expect to find new physics at the reach of the LHC.  

In this short review, focused on Higgs physics, we have not discussed the production of gravitational waves, that is a relevant feature of a SFOEPT (see, for instance Refs.~\cite{Huang:2016cjm,Carena:2019une,Vaskonen:2016yiu,Dorsch:2016nrg,Ellis:2020awk,Caprini:2019egz}). However, there tends to be tension between the bubble wall velocities associated with the very strong phase transition demanded for gravitational wave observations and the moderate ones necessary for proper diffusion of the baryon density
in electroweak baryogenesis scenarios. 

The requirement of new sources of CP-violation which couple to the Higgs also leads to the expectation of non-trivial
CP-violating Higgs couplings, inducing new collider signatures as well as electric dipole moments. Present constraints of 
these couplings imply that the tau-lepton modifications are the most likely to be associated with the generation of the
baryon asymmetry. Let us stress, however, that these conclusions are based on a simplified scenario where only the
third generation Higgs anomalous  couplings lead to the generation of the baryon asymmetry and to the electric dipole moment
contributions. If there is new physics at the weak scale, some of these conclusions may be modified.

Overall, the coming years will lead to a further probe of the Higgs sector and of new physics at the weak scale and hence of this exciting scenario for the generation of the matter - antimatter asymmetry.

~\\
~\\
{\bf \Large Acknowledgements} \\
~\\
We would like to thank H. Bahl, S. Baum, M. Carena, P. Huang, A.~Joglekar, G. Nardini, A. Megevand, A. Medina, A.~Menon, D. Morrissey, M. Quiros, A. Riotto,  N. Shah and Y. Wang  for many discussions and collaborations that led to the realization of this work.  C.W.  is supported in part by the U.S. Department of Energy grant DE-SC0013642. Work at Argonne is supported in part by the U.S. Department of Energy under contract DE-AC02-06CH11357.

\bibliographystyle{utphys}
\bibliography{HiggsBaryo}

\end{document}